# On the excitation of action potentials by protons and its potential implications for cholinergic transmission

Christian FILLAFER[1,*], Matthias F. SCHNEIDER[1,*]

[1]Biological Physics Group
Department of Mechanical Engineering
Boston University
110 Cummington St.
Boston, MA 02215
United States
phone: +001-617-353-3951

*To whom correspondence should be addressed: cf@bu.edu & mfs@bu.edu

## Abstract

One of the most conserved mechanisms for transmission of a nerve pulse across a synapse relies on acetylcholine. Ever since the Nobel-prize winning works of Dale and Loewi, it has been assumed that acetylcholine - subsequent to its action on a postsynaptic cell - is split into *inactive by-products* by acetylcholinesterase. Herein, this widespread assumption is falsified. Excitable cells (*Chara australis* internodes), which had previously been unresponsive to acetylcholine, became acetylcholine-sensitive in presence of acetylcholinesterase. The latter was evidenced by a striking difference in cell membrane depolarisation upon exposure to 10 mM intact acetylcholine ($\Delta V=-2\pm5$ mV) and its hydrolysate respectively ($\Delta V=81\pm19$ mV) for 60 sec. This pronounced depolarization, which also triggered action potentials, was clearly attributed to one of the hydrolysis products: acetic acid ($\Delta V=87\pm9$ mV at pH 4.0; choline ineffective in range 1-10 mM). In agreement with our findings, numerous studies in the literature have reported that acids excite gels, lipid membranes, plant cells, erythrocytes as well as neurons. Whether excitation of the postsynaptic cell in a cholinergic synapse is due to protons or due to intact acetylcholine is a most fundamental question that has not been addressed so far.

## 1. Introduction

The transmission of a nerve pulse from a presynaptic to a postsynaptic cell (*e.g.* nerve, muscle, secretory cell, etc.) is fundamental for many physiological functions. With the exception of a class of specialized synapses (gap junctions), this transmission process is considered to be of a chemical nature [1,2]. Many of our present ideas about chemical transmission have been influenced by works on cholinergic synapses (*i.e.* where acetylcholine (ACh) is the neurotransmitter) [3–7]. In short, the events taking place at such synapses are believed to be as follows: *(i)* A nerve pulse reaches the axon terminal and leads to the liberation of ACh. *(ii)* ACh translocates to the postsynaptic membrane and binds to a transmembrane protein (acetylcholine-receptor (AChR)). *(iii)* Acetylcholinesterase (AChE) hydrolyses and thereby deactivates ACh.

The enzyme AChE is an integral component of cholinergic synapses [8,9] where it can be anchored to parts of the basal lamina [10] and post- and presynaptic membranes [8,10]. Consensually, AChE has been attributed the role of „cleaner" of the synaptic cleft [2–7,9]. In contrast to this presently predominant conception, earlier models of the cholinergic synapse considered AChE to be the receptor for ACh or at least part of it [9,11]. Wurzel went a step further and suggested that choline ($Ch^+$), which is liberated in course of the enzymatically catalyzed hydrolysis of ACh,

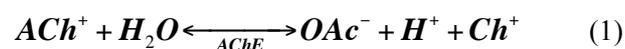

$$ACh^+ + H_2O \underset{AChE}{\longleftrightarrow} OAc^- + H^+ + Ch^+ \qquad (1)$$

could have excitatory effects [12,13]. At the same time any involvement of the simultaneously produced acetic acid, which dissociates into acetate (OAc$^-$) and protons (H$^+$)[*] respectively, was excluded based on pharmacological reasoning (see [13] and discussion therein; [8]). Nevertheless, Kaufmann proposed a model of the cholinergic synapse in which these hydrolytically liberated protons play a key role in excitation of the postsynaptic membrane [14–18]. Dettbarn and Hoekman came to a similar conclusion when interpreting experimental results with lobster axons [19,20]. It is the aim of the present work to test these possibilities. We will focus on the following hypothesis in the current theory of cholinergic transmission:

> *(A) Hydrolysis of ACh results in substances which - by themselves - do not have an appreciable effect on excitable membranes*

To the best of our knowledge, *no unambiguous evidence exists for this common assumption*. Control experiments, for instance, which have shown that acetate is inactive on muscle preparations [21], do not appropriately address the matter. Acetate is the conjugate base of acetic acid. As such, it can not mimic the acidification of the synaptic cleft which occurs in course of enzymatic hydrolysis of ACh (Eq. (1)). Even if the up-to-date postulated assumption *(A)* held true, adequate control experiments with hydrolysis products would be very valuable, because they can falsify alternative theories of cholinergic excitation such as those of Wurzel [12,13] and Kaufmann [14–18].

**Choice of model system.** Established cholinergic synapse models (*e.g.* frog neuromuscular junction, electric eel or *Torpedo* electroplaques, etc.) contain ACh-binding proteins [7,22] as well as AChE [8]. *Under physiological conditions* AChE activity is remarkably high in these synapses and results in quantitative hydrolysis of ACh [8,9] (Nachmansohn even went so far as to state that Loewi's experiments of detecting ACh extrasynaptically [3] can not be reproduced [9]). In most experiments on cholinergic synapses, however, AChE has been inhibited with anticholinesterases [8]. Such use of „specific pharmacological blockers" (*e.g.* physostigmine, diisopropyl fluorophosphate, α-bungarotoxin, etc.) in general – although it has been fruitful and central for the study of many problems in (neuro-) physiology [7,8,23] – entails potential pitfalls. To most readers this will be a moot point, but nevertheless it deserves notice that: *(i)* by adding foreign substances to the system one *creates and studies an unphysiological condition* [15,16] *(ii)* complete knock-down of AChE activity in the ~50 nm wide synaptic cleft and experimental confirmation of the latter is not trivial [16] *(iii)* attribution of the term „specific" to agonistic/antagonistic properties of a substance becomes increasingly difficult when one deals with multicomponent and –phase systems such as cells or

---

[*]throughout the manuscript the term „proton" will be used synonymously for its hydrated forms, *e.g.* H$_3$O$^+$, H$_5$O$_2^+$

tissue[†‡], *(iv)* any mistakes made when assigning such qualities as „specific" to a substance will be carried on in logical inferences[§].

In order to test assumption *(A)* and in an attempt to circumvent the above problems, we considered it to be advantageous to start from a system that is excitable (in the electrophysiological sense), but does not respond to ACh. Subsequently, one can introduce additional components (*e.g.* AChR, AChE, etc.) and one can observe if ACh-sensitivity *arises* by such manipulations. Hypothesis *(A)*, for instance, predicts that the simultaneous exposure of such a system to ACh and AChE will have no appreciable effect because the hydrolysis products are inactive. At the same time the current theory of cholinergic excitation requests that introduction of nicotinic or muscarinic AChRs will be *necessary* in order to confer responsiveness towards ACh upon a previously unresponsive excitable cell. From an evolutionary perspective one will be in the position to address the following question: *Which minimal set of components has to evolve at the interface of two excitable cells in order to allow for their communication through ACh?* In principle, a similar approach has been pursued with molecular biological techniques in established expression systems (*e.g. Xenopus* oocytes) [24]. These model systems are very valuable. Yet, it is important to recall that oocytes, for instance, possess *(i)* intrinsic ACh-sensitivity which has to be inhibited by atropine [25] and *(ii)* intrinsic AChE activity [24,26]. Herein, the applicability of excitable *Charophyte* cells for the purpose outlined above will be described. *Chara australis* internodal cells, which have a long standing history as model systems in electrophysiology [27–30], will be used to test hypothesis *(A)*. The effect of intact ACh on the membrane potential of these excitable cells will be characterized. Subsequently, the excitatory potency of reaction products of enzymatic hydrolysis of ACh by AChE – which are generated in cholinergic synapses and which have been considered to be inactive – will be investigated.

## 2. Materials and Methods

*Materials.* Acetylcholinesterase from *Electrophorus electricus* was obtained from Sigma-Aldrich (type V-S; Lot# 021M7025V). All other reagents were also purchased from Sigma-Aldrich (St. Louis, MO, USA) and were of analytical purity (≥99%).

***Cell cultivation and storage.*** *Chara australis* cells were cultivated in glass aquariums filled with a layer of 2-3 cm of New England forest soil, quartz sand and deionized water. The cells were grown under illumination from an aquarium light (14W, Flora Sun Max Plant Growth, Zoo Med Laboratories Inc., San Luis Obispo, CA, USA) at a 14:10 light:dark cycle at room temperature (~20°C). Prior to use, single internodal cells were stored for a minimum of 12 h in a solution containing 0.1 mM NaCl, 0.1 mM KCl and 0.1 mM $CaCl_2$ [30].

---

[†]from [7]: „*Initial attempts to identify the acetylcholine-binding site were hindered [because several effectors on electroplaque] also bind to [...] molecules distinct from the receptor and/or have high partition coefficients in lipidic compartments*"

[‡]from [76]: „*We have found that [α-bungarotoxin] has a high non-specific affinity for many substrates, especially glass and some plastics including Teflon*"

[§]The following *modus ponens* is an over-exaggerated example: If α-bungarotoxin binds with high affinity, the acetylcholine receptor is present. α-bungarotoxin binds to Teflon. Thus, the acetylcholine receptor must be present on Teflon.

*Experimental setup.* A single internodal cell (3-6 cm long) was placed on a plexiglass chamber into which compartments (~5 x 5 x 10 mm; h x w x l) had been milled. Small extracellular sections (length ~5 mm) of the cell were electrically isolated against each other with vacuum grease (Dow Corning Corporation, Midland, MI, USA). The $K^+$-anesthesia technique [29,30] in combination with extracellular electrodes (Dri-Ref, World Precision Instruments, Sarasota, FL, USA) was used for monitoring the cell membrane potential. Strictly, the membrane potential measured by this technique includes components from plasmalemma and tonoplast. It is expected that for extracellular addition of test solutions and for short timescales the contributions from the plasmalemma will dominate. The extracellular solutions contained 110 mM KCl in the first compartment and artificial pond water (APW) in all other compartments (0.1 mM KCl, 0.1 mM NaCl, 0.1 mM $CaCl_2$, 5 mM TRIS and 190 mM D-sorbitol; pH set to 7.0 with HCl). The potential between the virtual intracellular electrode (KCl-compartment) and an extracellular electrode was recorded with a voltage sensor (PS-2132; 50Hz sample rate; PASCO scientific, Roseville, CA, USA) and was defined as the membrane resting potential ($V_r$). All experiments were conducted at room temperature (~20±2 °C).

*Test solutions.* APW was acidified with HCl and acetic acid respectively to a desired pH (monitored with a glass electrode; Van London–pHoenix Co., Houston, TX, USA). Solutions of acetylcholine chloride (10, 25, 50 mM) and choline chloride (1, 5, 10 mM) were prepared fresh prior to each experiment. For these solutions, APW was used as described above except that the amount of D-sorbitol was decreased to ensure a constant osmolarity of 200±2 mOsmol $kg^{-1}$ (3D3 Osmometer, Advanced Instruments, Norwood, MA, USA). Hydrolysis of ACh (10 mM) was carried out in 20 mL APW (composition as described above; instead of 5 mM TRIS, only about 1-2 mg were added to buffer the drop in pH that had resulted from spontaneous hydrolysis of ACh). This solution was stirred in a glass beaker and after about 10 min AChE was added (5 μL of a stock solution in APW (~500 units per 100 μL)). At fixed points in time (corresponding to a, b, c, d, e in Fig. 2) samples were drawn with a micropipette.

*Effect of test solutions on membrane potential.* APW was removed from one of the compartments and subsequently this section of the cell was exposed to a test solution for 60 sec (Fig. 1, 2, 3) and ~2 sec respectively (Fig. 4). The resting membrane potential ($V_r$ ~ -200 mV) was subtracted from the value of the membrane potential after 60 sec ($V_m$). Thus, positive values of $\Delta V=V_m-V_r$ indicate depolarization while negative values indicate hyperpolarization. Experiments in which an AP was triggered were not included in the calculation of averages of ΔV. Typically, a single cell was exposed to 6-8 test solutions with two washing steps (with APW) and a recovery period (15 min) in between trials. The sequence of exposure of a single cell to different test solutions was varied arbitrarily and was not found to have a biasing effect on the results obtained.

## 3. Results and Discussion

*ACh and the Chara membrane potential.* In general, limited knowledge is available about the occurence of AChR, AChE as well as ACh in Charophyte cells (*e.g. Nitella*, *Chara*, etc.). AChE activity has been detected in *Nitella flexilis* homogenates, but the cellular location of the enzyme was not clarified [31]. Patch-clamp studies with tonoplast membrane from *Chara corallina* indicated a certain responsiveness to ACh (4-6mM) [32]. It has also been reported that the repolarization phase of an electrically triggered action potential (AP) is prolonged upon extracellular addition of >1mM ACh in *Chara corallina* [32] and *Nittelopsis obtusa* [33]. None of these studies, however, have found excitation of the cell by extracellular addition of ACh. In the present work, we were mainly concerned with quick (timescales <60 sec) effects of ACh on the membrane potential ($V_m$) of *Chara* internodal cells. As illustrated in Fig. 1, a minor depolarization (ΔV=12±6 mV), which typically had an irregular fluctuating timecourse, was detected upon extracellular addition of 50 mM ACh for 60 sec. Most importantly, however, at concentrations of ACh <50 mM no sign of excitation in the sense of *(i)* pronounced depolarization or *(ii)* triggering of an AP was observed. Thus, *Chara australis* cells were considered to be a suitable model system for studying the problems discussed above.

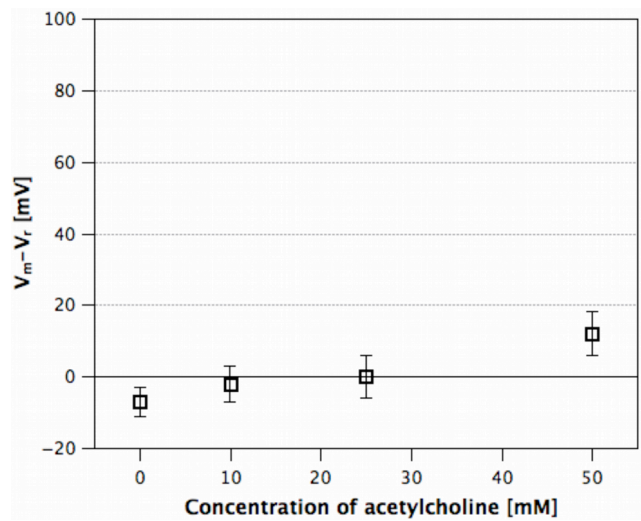

**Figure 1. Effect of acetylcholine on membrane potential of *Chara australis* cells.** Difference between resting potential ($V_r$; typically about -200 mV) and potential after exposure to ACh for 60 sec ($V_m$). Data points are averages of n=3-6 cells ± StDev.

***Hydrolysis products of ACh are excitatory.*** In the next step, the consequences of introducing AChE to this system were tested. The rationale of the experiment was as follows: One starts in a beaker with an aqueous solution of ACh that contains the highest possible amount of intact ACh[**]. Upon addition of AChE, the concentration of ACh will decrease and simultaneously, hydrolysis products (AcH, OAc$^-$, H$^+$, Ch$^+$) will accumulate. This process represents the ubiquitously occuring degradation of ACh in cholinergic synapses. The evolving reaction can be followed easily by use of a pH electrode (Fig. 2, *top left*). The amount of intact ACh in samples (*a, b, c, d, e*), which are drawn from the beaker at acetic acid as monitored by pH measurements. *(top right)* Effect of hydrolysate samples (**a**, **b**, **c**, **d**, **e**) on membrane potential of *Chara australis* cells within 60 sec. Positive values of $V_m$-$V_r$ indicate depolarization. Data points are averages of n=6-14 measurements ± StDev. *(bottom)* Individual cell membrane potential traces upon addition (grey bar) of respective sample. Triggering of action potentials (arrows). Scale bar represents 10 sec.

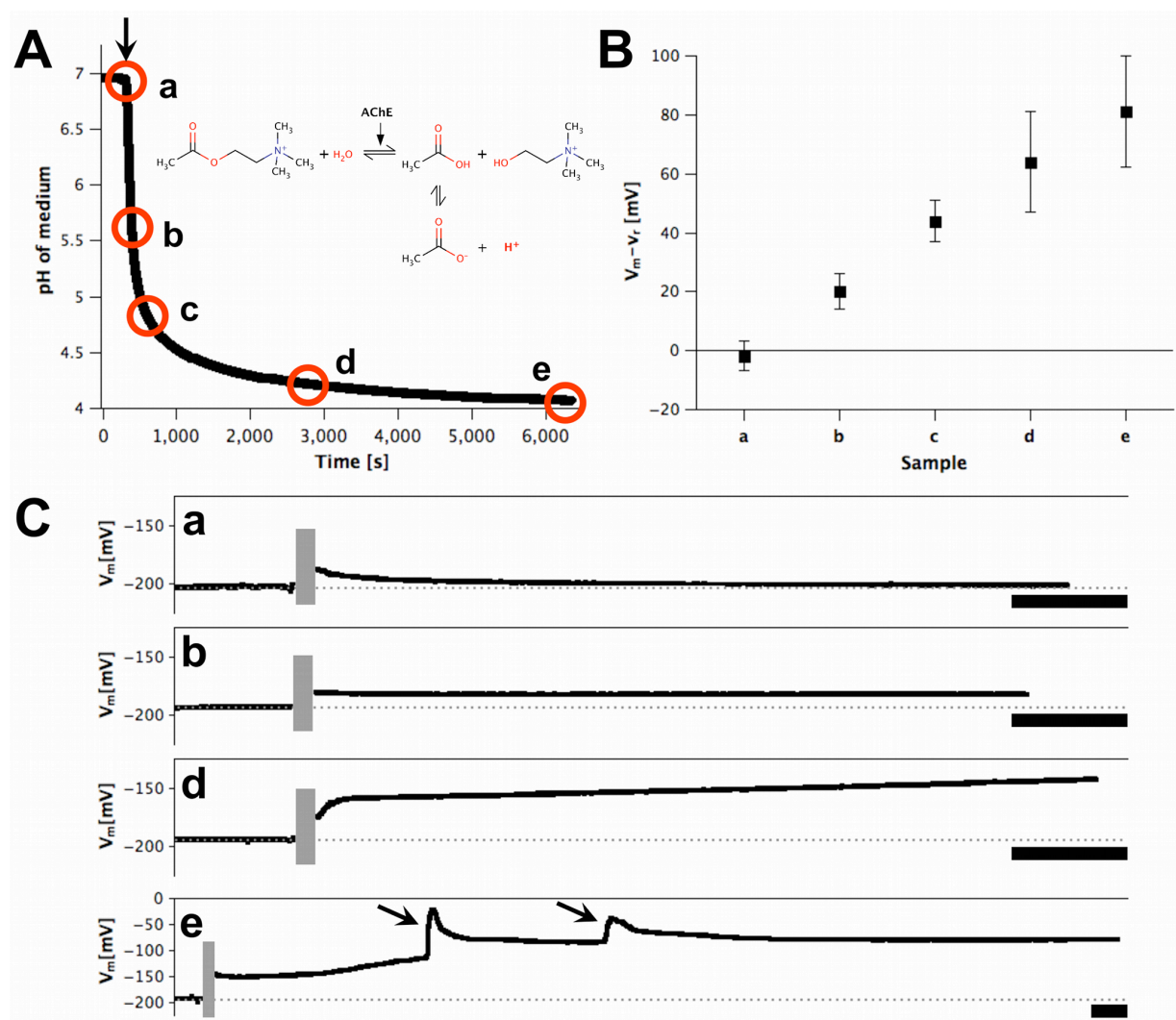

**Figure 2. Reaction products of enzymatically catalyzed hydrolysis of ACh are excitatory**. *(top left)* Hydrolysis of ACh is accelerated by AchE (addition indicated by arrow) and liberates choline and

---

[**]it is not possible to apply pure ACh since the compound is hydrolyzed spontaneously, yet at a slow rate, in aqueous solution

progressive stages of the reaction, will obey the relation $[ACh]_a > [ACh]_b > ... > [ACh]_e$. If hypothesis *(A)* were correct, one would expect that degradation of ACh by AChE will lead to particulary „inactive" solutions. Thus, in general the following excitatory potency is predicted: **a** > **b** > **c** > **d** > **e**. In the present case none of the hydrolysates should have an effect on the excitable cell (compare Fig. 1 and hypothesis **(A)**). *The experimentally observed depolarization of the Chara cell membrane upon exposure to the hydrolysates, however, is in stark contrast to this prediction* (Fig. 2, *top right*). The more ACh is degraded, the higher becomes the excitatory potency of the test solution. This was clearly evidenced by the degree of depolarization of the membrane and by the increased likelihood of triggering of action potentials (Fig. 2, *bottom*; in 4 out of 10 cases at pH ~4.0). The common assumption *(A)* is thus considered to be falsified.

*Identity of the excitatory agent.* It is of interest to identify if the observed excitatory effect has to be attributed to *(i)* choline (suggested by Wurzel [12,13]) or *(ii)* acid (suggested by Kaufmann [14–16]). As illustrated in Fig. 3 *(left)*, choline up to concentrations of 10mM had no detectable effect on the

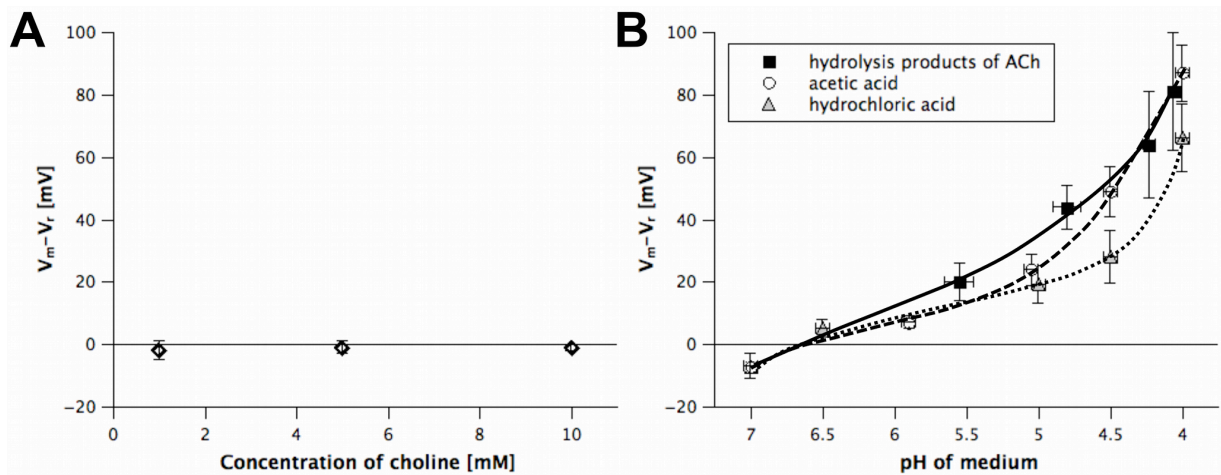

**Figure 3. Effects of choline, acetic- and hydrochloric acid on membrane potential of *Chara australis* cells.** Positive values of $V_m$-$V_r$ indicate depolarization within 60 sec. Data points are averages of n=3 (choline), n=6-9 (acetic acid), and n=9-11 measurements (hydrochloric acid) ± StDev. Lines are guides to the eye. The results presented above are corroborated by qualitatively identical observations of acid-induced depolarization in other *Charophyte* species (20).

*Chara* cell membrane potential. In contrast, artificial pond water that had been acidified with acetic acid closely reproduced the depolarization that was observed with the hydrolysates (Fig. 3, *right*). Moreover, action potentials were triggered with acetic acid-containing test solutions of pH 4.0 (in 2 of 7 cases). Further experiments were conducted with artificial pond water that had been acidified with hydrochloric acid. These studies clearly indicated that the overall phenomenology is invariant to the type of acid employed[††] (Fig. 3, *right*). It was also found that application of test medium acidified to pH ≤3.5±0.2 with HCl resulted in immediate and repeatable triggering of APs within 1-2 sec (Fig. 4).

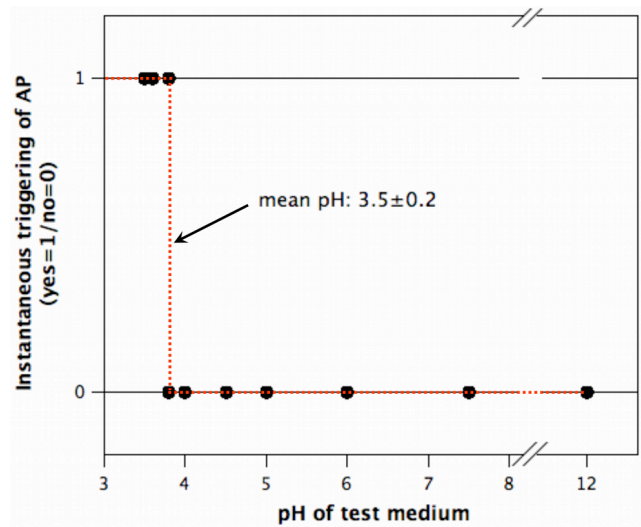

**Figure 4.** Instantaneous triggering (within ~2 sec) of APs in a *Chara australis* cell upon exposure to acidic (HCl) but not neutral or alkaline test media (NaOH) respectively. Mean pH-threshold in n=4 cells ± StDev.: 3.5±0.2

This indicates that the timescales for induction of an AP are strongly dependent on proton concentration. *It is concluded from these results that protons are the main excitatory component of hydrolysates of ACh.*

***Protons excite neurons, gels and lipid membranes.*** The sensitivity of cell membranes to protonation is by no means a peculiarity of plant cells. In fact, the typically negatively charged cell surfaces of eukaryotes [34] make it seem unavoidable that a cell is susceptible to pH. A proton-induced increase of membrane conductance has indeed been shown to occur in neurons from trigeminal ganglia [35,36] and neuroblastoma cells [36]. Membrane depolarization and/or triggering of action potentials have been reported in cortical neurons of cats [37], cultured spinal neurons from mice [38], lobster giant axons [19,20] as well as *Xenopus laevis* oocytes [39]. It is also possible that effects of acids have been demonstrated *unintentionally* in experiments involving iontophoretic application of neurotransmitters.

---

[††] it is probable that the excitatory potency of an acid varies based on *e.g.* its pK, solubility, etc.

Frederickson et al., for instance, showed that protons are expelled preferentially from norepinephrine-containing micropipettes due to their higher electrophoretic mobility [37]. The very same issue might have biased studies on cholinergic synapses. There, ACh has frequently been applied from micropipettes containing a 1M solution of ACh [40,41]. Within seconds upon preparation, such a solution acquires a pH of ~3.25 due to spontaneous hydrolysis of ACh (observations in our laboratory). Thus, it is conceivable that the iontophoretic application of ACh in *e.g.* [40,41] was to some degree a iontophoretic application of protons. Further evidence for effects of acidic pH on neurons has been reviewed recently [42–44]. It is also well-established that model systems of biological membranes such as lipid mono- and bilayers [45–47] as well as gels [48–50] are affected by pH changes. Träuble, for instance, has demonstrated theoretically and experimentally that one of the thermodynamic properties of a membrane is its pK and that titration with acid or base is a potent means to induce thermodynamic state changes [45,46]. Kaufmann and co-authors have shown that protonation-induced state changes can lead to the induction of transmembrane current fluctuations in protein-free lipid membranes[‡‡] [51–53]. It has also been reported that rapid application of acid droplets onto a charged lipid monolayer leads to excitation of pulses that propagate over macroscopic distances [54]. Another beautiful example for protonation-induced state changes are Kuhn's artificial muscle models, which exhibit reversible macroscopic shortening and extension upon acidification and alkalinization respectively [49,50].

***Bulk versus interfacial protonation.*** In most experiments aimed at understanding the effects of pH on cells and tissue, protons are typically applied via the bath or at considerable distances from the excitable membrane. This „bulk" approach, which also has been pursued in the present study, seems rather crude in comparison to the refined mechanisms that have evolved in nature. In biological systems, the majority of water exists at hydrated interfaces (membranes, polymers (such as cytoskeletal filaments or DNA), etc.). Enzymes located at such an interface (*e.g.* AChE, carbonic anhydrase, ATPase, etc.) have the potential to rapidly switch its thermodynamic state by catalytic activity. This is best illustrated in the system which has been addressed in the present work: the cholinergic synapse. The latter can be considered a hydrolytic „hotspot" for ACh due to the remarkably high local acetylcholinesterase activity (8). Under physiological conditions (= no anticholinesterase present), a considerable fraction [55–57] or almost all of the ACh [15,16] liberated in a synapse will be rapidly degraded before it has the opportunity to interact with other components (*e.g.* an ACh-binding protein AChR). An order of magnitude estimate based on data from rat diaphragm[§§] [58,59] indicates typical timescales for hydrolysis of most of ACh within ~1-2 msec. If

---

[‡‡] the interested reader is referred to evidence in [70–72,77,78] that contradicts the concept of ion translocation through protein channels

[§§] area of endplate: ~7000 $\mu m^2$; volume of endplate: ~450 $\mu m^3$; catalytic site density: ~2500 sites $\mu m^{-2}$; hydrolysis rate: 0.3 molecules catalytic site$^{-1}$ msec$^{-1}$; ACh release: $3 \cdot 10^6$ molecules pulse$^{-1}$ endplate$^{-1}$ (the latter value from [59] has been quoted frequently despite the fact that the calculation and assumptions taken were not

one treats the synaptic cleft as a bulk volume into which reaction products are released, this corresponds to a drop of pH from ~7 to ~4. However, such an analogy unlikely captures the conditions in the synapse. It is conceivable that protons, instead of being released into an aqueous bulk phase, are directly buffered by the negatively charged hydrated interfaces into which AChE is embedded. Such an AChE-induced increase in local proton concentration has indeed been reported to occur on artificial collodion membranes [60], *Electrophus electricus* membrane fragments [61], and „modified" electroplaque cells [62]. Moreover, it has been shown [52,53,63–65] that incorporation of AChE in lipid bilayers renders the membrane sensitive for ACh (which was attributed to local protonation of the interface [52,53]). Cell membranes of erythrocytes from several species (human, guinea pig, dog, etc.) also contain AChE. When exposed to ACh, these cells are progressively permeabilized due to enzymatically catalyzed acidification [66]. Although biological effects due to interfacial protonation by AChE seem to be rather common, they have not been considered in quantitative models of the cholinergic synapse [55–57] nor in the respective literature (*e.g.* [6,7,43]).

***Some remarks on cholinergic transmission and the nature of receptors.*** The current theory of cholinergic transmission has been elaborated in remarkable electrophysiological and molecular detail [7]. Nevertheless, it is of utmost importance to challenge its underlying assumptions. Herein, it was demonstrated that intact ACh does not particularly affect an excitable plant cell. However, catalytic hydrolysis of ACh generates protons which depolarize cell membranes and excite action potentials (*i.e.* hypothesis *(A)* was falsified). Further evidence corroborating this view has been reported in the literature [19,20,35–38,42–44].

Based on these findings, we suggest that our current conceptions of cholinergic transmission need to be revisited. To facilitate the latter, we propose the following falsifiable hypothesis: Excitation in cholinergic synapses is the result of an adiabatic state change of the postsynaptic membrane induced by protonation (as has been elaborated on by Kaufmann [14–18]). Some consequences implied by this hypothesis are: *(i)* The sufficient components for cholinergic transmission are ACh, AChE and a charged hydrated interface (this argument is based on parsimony[***] and does not preclude the

---

detailed in the paper. Moreover, it is not clear if residual AChE activity (after eserin treatment) was accounted for. Generously assuming 99% of inhibition, this still leaves a hydrolysis rate of ~$5 \cdot 10^4$ molecules msec$^{-1}$ endplate$^{-1}$. In [59], ACh release was determined in the bath ~300 sec after repetitive stimulation. During this timespan ~$1.5 \cdot 10^{10}$ additional molecules of ACh could have been hydrolyzed per endplate and would not appear in a detection essay. Thus, the actual concentration of ACh per pulse per endplate could easily have been $3 \cdot 10^7$ molecules – an order of magnitude larger than assumed – or even higher (as argued by some: 0.1-1mM; see discussion in 11)). Hydrolysis of ~$10^7$ ACh molecules will thus take ~1-2 msec under physiological conditions. In the presence of cholinesterase inhibitors the reaction is not stopped but simple extended in time to ~10-100 msec. These order of magnitude estimates are in good agreement with experimentally obtained timescales of excitatory postsynaptic potentials (compare Fig. 6 in [5]).

[***] from I. Newton's Rules of Reasoning in Philosophy: „*We are to admit no more causes of natural things than such as are both true and sufficient to explain their appearances. Therefore, to the same natural effects we must, so far as possible, assign the same causes*"

realization of more complicated systems[†††]). *(ii)* In the current cholinergic theory acetylcholinesterase performs work „against the receptor" by rapidly degrading its ligand. The inherent inefficiency of this kinetic competition would be resolved if one assumed that AChE generates the excitatory agent [16]. *(iii)* The postsynaptic membrane is specific for acetylcholine because of the presence of highly specific enzymatic activity. *(iv)* Currently and conventionally the term „receptor" is used for a specific molecule that exerts a function upon binding of a ligand [7,67]. We would like to shortly elaborate on an alternative to this molecular viewpoint which has been motivated by Einstein's approach to thermodynamics [68,69]: As has been documented above, simple (pure lipid membranes, gels, etc.) as well as compositionally complex hydrated interfaces (biological membranes) – systems with entirely different molecular composition – are highly susceptible to pH. *Thus, it seems intuitive to consider the entire system (e.g. the cellular membrane including lipids, proteins, carbohydrates, attached extracellular matrix, hydration layers, etc.) and not a single molecular entity as „the receptor"*. Such a *receptive system* reveals its functions in a particularly impressive manner upon drastic changes in its thermodynamic state, *i.e.* near transitions. Therefore, we propose that the thermodynamic state diagrams of a system determine its ability to receive stimuli and to exert functions (state-function-relationship). Protonation[‡‡‡] of a cell membrane (receptive system), for example, will induce a state change. In particular when near a transition, this new state of the system will manifest dramatically changed biological functions (*e.g.* permeability [51,70–72], affinity for ions (*e.g.* $Ca^{2+}$) and catalytic activity [73–75]). Yet, neither reception nor function are created by single molecules in an inert matrix. They arise as inseparable consequences of the receptor being a thermodynamic system which experiences state changes.

## Acknowledgements

We thank **K. Kaufmann** for advocating the importance of interfaces in biology and for introducing the analogy of an enzymatic "proton pistol". Moreover, we are thankful for his stimulating lectures and the ensuing discussions on the thermodynamics of soft interfaces. I. Silman, B. Fichtl, S. Shrivastava, H. Kong and W. Hanke have provided helpful criticism of the manuscript. *Chara* cells for starting our cultures were kind gifts of W. Hanke, I. Foissner, M. Bisson and R. Wayne. We also thank D. Campbell for crafting measuring chambers.

---

[†††] for instance, butyrylcholinesterase or spontaneous hydrolysis could contribute to the liberation of protons from ACh [79]

[‡‡‡] in fact, any variation of a thermodynamic variable (temperature, dissolution of *e.g.* ethanol, change of ion concentrations, mechanical extension, etc.) to which the system is susceptible